\documentclass[%
 reprint,
 amsmath,amssymb,
 aps,
]{revtex4-2}

\usepackage{graphicx}
\usepackage{dcolumn}
\usepackage{bm}

\usepackage[linesnumbered,ruled,vlined]{algorithm2e}
\usepackage{algcompatible}
\usepackage{subcaption}
\usepackage{yfonts}

\newcommand{\A}{{\mathcal A}}
\newcommand{\U}{{\mathcal U}}
\newcommand{\Q}{{\mathcal Q}}
\newcommand{\K}{{\mathcal K}}
\newcommand{\I}{{\mathcal I}}
\newcommand{\G}{{\mathcal G}}
\newcommand{\R}{{\mathcal R}}
\newcommand{\M}{{\mathcal M}}

\newcommand{\T}{{\mathcal T}}
\newcommand{\V}{{\mathcal V}}

\begin{document}

\preprint{APS/123-QED}

\title{Quantum error mitigation for parametric circuits}

\author{Vasily Sazonov}
\email{vasily.sazonov@cea.fr}
\author{Mohamed Tamaazousti}%
\email{mohamed.tamaazousti@cea.fr}
\affiliation{%
 Universit\'e  Paris-Saclay,  CEA,  List,\\  \textit{F-91120,  Palaiseau,  France}}%

\date{\today}

\begin{abstract}
Reducing errors is critical to the application of modern quantum computers. In the current Letter, we investigate the quantum error mitigation considering parametric circuits accessible by classical computations in some range of their parameters. We present a method of global randomized error cancellations (GREC) mitigating quantum errors by summing a linear combination of additionally randomized quantum circuits with weights determined by comparison with the referent classically computable data. We illustrate the performance of this method on the example of the $n = 4$ spins anti-ferromagnetic Ising model in the transverse field.
\end{abstract}

\maketitle

\section{Introduction}
The unavoidable coupling of quantum devices to the environment results in noise in quantum channels and gates and limits practical applications of quantum computations. Although the general theory of quantum error corrections provides a path for implementing fully fault-tolerant computations \cite{ShorCode, Steane, Calderbank, ChengBook}, the needed qubits overheads are insurmountable for modern quantum computers \cite{Jones, Arute2019, IBM}.

As an alternative to the quantum error correction, one may consider a quantum error mitigation approach \cite{Mitig, PractMitig}. The main challenge for the error mitigation algorithms is to achieve a reasonable performance for large-scale circuits. Among the most important factors limiting the development and application of error mitigation, one can distinguish the lack of information regarding the the noise properties and the absence of the relevant data to benchmark and correct results of computations \cite{LearnMitig, NNMitig}. In the current work, we eliminate the problem of data deficiency by focusing on quantum simulations of physical systems for which classical computations are accessible for some regions of their parameters and tackle the characterization of the noise by running auxiliary quantum computations.

Indeed, there are many physical systems having regimes with different classical computational complexities. These regimes can be 'easy' where, for instance, the Monte Carlo methods \cite{Metropolis, Hastings, Barker} work in polynomial time and 'hard' where effective classical solutions are not known.

\begin{figure}[h]
\centering
\includegraphics[width=0.45\textwidth]{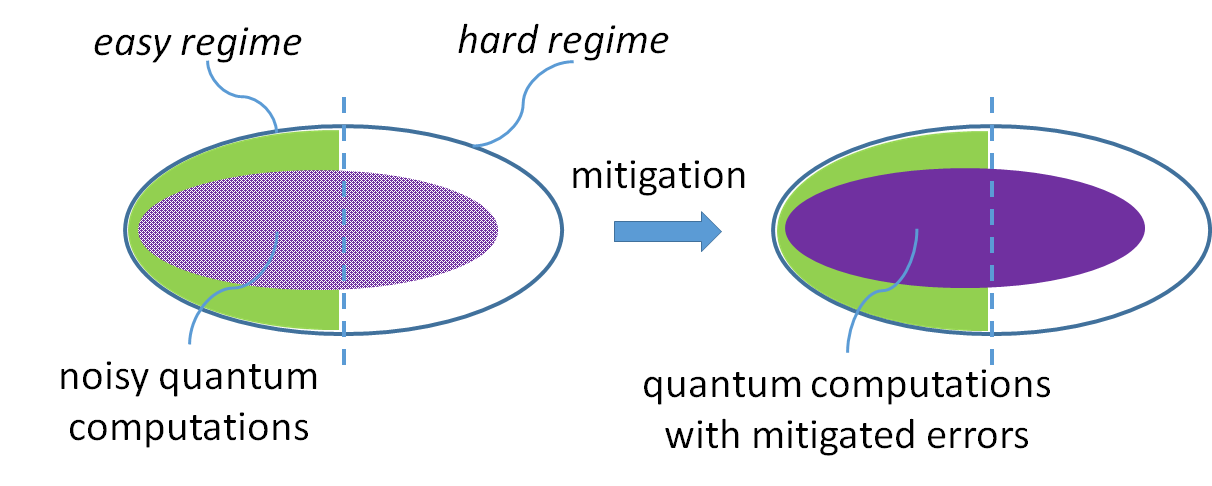}
\caption{\small The proposed setup for learning the error mitigation.}
\label{fig:scheme}
\end{figure}

For example, in lattice Quantum Chromodynamics (QCD) at non-zero chemical potentials, the failure of the Monte Carlo methods is caused by the infamous \textit{sign problem} \cite{Forcrand2009, Gattringer, SPel}.
This problem prevents classical simulations of the quark matter at non-zero densities relevant e.g., for the description of the interior of the neutron stars \cite{Orsaria} and heavy-ion collisions \cite{heavyIon}. At the same time, the \textit{sign problem} in lattice QCD is absent at the nonphysical purely imaginary chemical potentials \cite{Forcrand2009}.
A similar situation holds for the Hubbard model: it is classically tractable at imaginary chemical potentials and possesses the \textit{sign problem} for real chemical potentials away from half-filling, where it is believed to describe the high-temperature superconductivity phase transition \cite{ImHubbard}.
Simulations on the quantum computer are free from the sign problem \cite{QAlgSP} and, if not the noise, they could be the primary tool for studying quantum many-body systems. In this Letter, we propose a new setup for quantum error mitigation dedicated to such context. Within our approach, the error mitigation is learned in 'easy' (e.g., \textit{sign problem}-free) regimes and applied in 'hard' ones, see the schematic representation of such a process in Fig. \ref{fig:scheme}. The green area corresponds to an 'easy' regime, where the classical computations are efficient, the white area represents a 'hard' one from the classical point of view. Noisy quantum computations -- light purple area -- are assumed to be feasible in both regimes. The overlap of green and light purple areas is used for training the error mitigation resulting in the error mitigated improved quantum computations valid for 'easy' and 'hard' classical regimes -- dark purple area.

In the following, we start with a general discussion on quantum error mitigation. Then, we develop the concept of learning the error mitigation in 'easy' regimes and formulate a global randomized error cancellation (GREC) algorithm, decreasing quantum errors by assembling additionally randomized quantum circuits.
We derive estimates on the performance of GREC based on the latest achievements in the theory of extrapolations \cite{Stable}.
Then, we benchmark the GREC algorithm on the $n=4$ spins anti-ferromagnetic Ising model in the transverse field carrying computations on the quantum simulator and quantum processor \textit{ibmq\_manila} provided by the IBM quantum experience \cite{IBMQ}. We conclude by discussing the most promising future applications of GREC.

\section{Error mitigation}
We start by discussing general aspects of error mitigation to relate our developments to the existing methods, such as zero noise extrapolation (ZNE) and probabilistic error cancellation (PEC) \cite{Mitig}.

Consider the quantum circuit $\U(\lambda)$ depending on the parameter $\lambda \in \Lambda \subset \mathbf R$ composed by the $k$ local unitary gates,
\begin{equation}\label{eq:u-circuit}
\mathcal U(\lambda) = \Gamma_k \circ \dots \Gamma_2 \circ \Gamma_1.
\end{equation}
The expectation value of an observable $\A$ measured after applying this circuit to $N_\Q$ qubits initially prepared in the state $\rho_0=|0\rangle \langle 0 |^{\otimes N_\Q}$ is given by
\begin{equation}\label{eq:a-expval}
\langle \A \rangle^{\rm ideal} = {\rm tr}[\A \hat\U(\rho_0)]\,,
\end{equation}
where $\hat\U(\rho_0) = \U \mathcal \rho_0 U^\dag$.
Usually, due to the hardware limitations and noise, the ideal gates $\Gamma_i$ cannot be implemented, and one is restricted only to a set of noisy operations $\mathcal O \in \mathcal I$.
However, if a set of noisy operations $\{\mathcal O_{\alpha} \} \subset \I$ forms a complete basis, any ideal gate $\Gamma_i$ can be represented as a linear combination of these operations:
\begin{equation}\label{eq:qpr}
\Gamma_i = \sum_\alpha \eta_{i, \alpha} \mathcal O_{i, \alpha}, \quad \mathcal O_{i, \alpha} \in \mathcal I, \quad  \eta_{i, \alpha} \in \mathbf{R}
\,,
\end{equation}
with the coefficients $\eta_{i, \alpha}$ which can be negative. Consequently, by means of equations \eqref{eq:u-circuit}, \eqref{eq:a-expval} and \eqref{eq:qpr} the ideal expectation value $\langle \A \rangle^{\rm ideal}$ can be expressed as a linear combination of many noisy expectation values:
\begin{equation}
\langle \A \rangle^{\rm ideal}= \sum_{\vec{\alpha}} \eta_{\vec{\alpha}} \langle \A_{\vec{\alpha}}\rangle^{\rm noisy}\,,~~\eta_{\vec{\alpha}} := \eta_{k, \alpha_k} \dots \eta_{2, \alpha_2} \eta_{1,\alpha_1}\,,
\label{eq:a-ideal-from-noisy}
\end{equation}
where coefficients $\eta_{\vec{\alpha}}$ form a quasi-probability distribution \cite{Mitig, PECpZN} and
\begin{align}
\langle \A_{\vec{\alpha}}\rangle^{\rm noisy} &:=  {\rm tr}[\A \Q_{\vec{\alpha}}(\rho_0)]\,,\\
\Q_{\vec{\alpha}} &:= \mathcal O_{k, \alpha_k} \circ \dots \circ \mathcal O_{2, \alpha_2} \circ \mathcal O_{1, \alpha_1}\,. \label{eq:noisy-circuit}
\end{align}
When the decomposition \eqref{eq:qpr} of the elementary gates $\Gamma_i$ is known and when the coefficients $\eta_{\vec{\alpha}}$ are mostly of the same sign, the PEC methods, employing the Monte Carlo evaluation of the sum in \eqref{eq:a-ideal-from-noisy}, can be applied for the error mitigation \cite{Mitig, PECpZN}. Alternatively, one can construct a representation similar to \eqref{eq:a-ideal-from-noisy} from scratch,
\begin{equation}
\langle \A \rangle^{\rm ideal}= \sum_{i} \eta_{i} \langle \A\rangle^{\rm noisy}_{i}\,,
\label{decomp}
\end{equation}
applying the ZNE approach \cite{Mitig, PECpZN}, which does not require any \textit{a priori} knowledge about decomposition's \eqref{eq:qpr}. 

In the following, we focus on constructing the error mitigation scheme agnostic to the elementary gates decomposition's \eqref{eq:qpr} but benefiting from the existence of the training data accessible in the classically tractable regimes of quantum circuits. The proposed GREC error mitigation algorithm is targeted for obtaining the ZNE-type representation of an ideal expectation of the observable \eqref{decomp}. The additionally randomized quantum circuits used by GREC are obtained by the modifications of elementary gates of the original circuit and are ideologically similar to the terms of the sum \eqref{eq:a-ideal-from-noisy}. However, in opposition to the PEC methods, the GREC mitigation algorithm determines the importance of such terms not by the absolute values of their amplitudes but according to the accessible training data.

\subsection{Global randomized error cancellation (GREC)}
Let us begin with necessary notations regarding the parametric dependence of quantum circuits resulting in regimes with different classical computational complexities.
To be more precise in definitions of these regimes, we suppose that the expectation value $\langle \A \rangle^{\rm noisy}$ can be evaluated on the quantum computer for all $\lambda \in \Lambda$ and effectively calculated on the classical computer when $\lambda \in \K \subset \Lambda$, as a concrete example of such sets one can consider $\K = [-1, 1]$ and $\Lambda = \mathbf R$. Let $\{\A\}^{\rm exact} = \{\langle \A(\lambda_1) \rangle,\cdots,\langle \A(\lambda_{N_\K}) \rangle\}$ be the exact values obtained at $\{\K\} = \{\lambda_1, \cdots, \lambda_{N_\K}\} \subset \K$ and $\{\A\}^{\rm noisy} = \{\langle \A(\lambda_1) \rangle^{\rm noisy},\cdots,\langle \A(\lambda_{N_{\Lambda}}) \rangle^{\rm noisy}\}$ be the values obtained from the noisy circuit at points $\{\Lambda\} = \{\lambda_1, \cdots, \lambda_{N_\K}\} \cup \{\lambda_{N_\K + 1}, \cdots, \lambda_{N_\Lambda}\}$. Due to the assumed dependence only on the one dimensional parameter $\lambda$, we refer $\{\A\}^{\rm exact}$ and $\{\A\}^{\rm noisy}$ as the exact and noisy curves respectively.

\subsubsection{Baseline}
The simplest version of the GREC algorithm, not involving any randomizations, can be formulated as a two-parameter fitting procedure approximating an ideal value of an observable as
\begin{equation}
  \langle \A \rangle^{\rm ideal} \approx \eta_1 \langle \A_{0}\rangle^{\rm noisy} + \eta_0\,.
\label{two_param}
\end{equation}
The latter representation can be justified for each $\lambda$ separately for the depolarised noisy channel model \cite{CliffordData}. The coefficients $\eta_0, \eta_1 \in \mathbf R$ are the subject for the linear regression on the overlap of the sets $\{\A\}^{\rm exact}$ and $\{\A\}^{\rm noisy}$ with $\lambda \in \K$.

The ansatz \eqref{two_param} can be efficient only in case of the weak dependence of quantum errors on $\lambda$. In reality, the errors from the individual gates can differently propagate through the circuit at different values of the parameter $\lambda$, which can lead to the significant dependence of the final result on the accessible region $\K$.

\subsubsection{Main algorithm}
\label{sec:GREC}
For constructing an error mitigation scheme universal for all $\lambda \in \Lambda$ we use additional quantum computations providing the way to incorporate the dependence on $\lambda$ implicitly and focus on finding the 'near optimal' expansion of the type \eqref{decomp}. 
To achieve the desired representation, we propose the following algorithm, where the auxiliary information characterizing the noise nature is obtained by running additionally artificially randomized quantum circuits.

\begin{algorithm}[H]
\KwIn{quantum circuit $\U(\lambda)$, $\{\K\}, \{\Lambda\}, \{\A\}^{\rm exact}$,\\
~~~~~~~~~~$N_\R$ - number of randomized circuits,\\
~~~~~~~~~~$N_\G$ - number of additional gates,\\
}
\KwOut{$\{\A\}_\Lambda^{\rm mitigated}$ - an observable with mitigated quantum errors}
Generate $N_\R$ additionally randomized circuits, each by: \linebreak
  (a) randomly inserting $N_\G$ gates $\G_l(\vec\theta_l)$ in circuit $\U(\lambda)$.
  \linebreak
  (b) preparing a random configuration of uniformly distributed 
  parameters $\theta_l^p \in [-\Delta, \Delta]$ with $l = 1..N_\G$, $p = 1..{\rm dim}(\vec\theta_l)$. $\Delta = 0$ corresponds to the original circuit $\U(\lambda)$.\\
Measure the randomized noisy curves $\{\A\}_\Lambda^{(r)} = \{\langle \A(\lambda_1) \rangle^{(r)},\cdots,\langle \A(\lambda_{N_{\Lambda}}) \rangle^{(r)}\}$, $r = 1..N_\R$ \footnote{In principle, one can include a non-randomized measurements ${\A}_\Lambda \equiv {\A}_\Lambda^{(0)}$ into this set, however we didn't find any remarkable advantage of it. }.\\
Minimize
\begin{equation}
  \sum_{\lambda \in \K} \Big(\sum_{r = 1}^{N_\R} \eta_r \langle \A(\lambda) \rangle^{(r)}  + \eta_0 -  \langle \A(\lambda) \rangle\Big)^2\,,
\label{opt}
\end{equation}
subjected to $\sum_{r = 1}^{N_\R} \eta_r = 1$ \footnote{This constrain is introduced to make the derivations of Section \ref{sec:stab} more transparent, and it was omitted in the baseline version of the GREC algorithm.}.\\
Construct an observable with mitigated quantum errors as
\begin{equation}
  \{\A\}_\Lambda^{\rm mitigated} = \sum_{r = 1}^{N_\R} \eta_r \{\A\}_\Lambda^{(r)}  + \eta_0\,.
\label{constr}
\end{equation}
\caption{GREC}\label{alg:GREC}
\end{algorithm}

If the number of the randomized curves $N_\R = 0$ and also the number of auxiliary gates $N_\G = 0$ we return to the baseline version of algorithms additionally constrained by $\eta_1 = 1$.

We note that the general procedure of the randomization of an initial quantum circuit can be modified in several ways. For instance, instead of placing auxiliary gates at random positions in the original circuit, one can consider these positions to be fixed and variate only the values of the auxiliary gates parameters. The types of gates $\G_l(\vec\theta_l)$ can be rather general and one can choose to work with gates acting on one, two, or more qubits. In our numerical experiments presented in Section \ref{ExRes}, we chose the strategy of equipping each different single gate in the original circuit by the most general one-qubit $U(\vec\theta)$ gate.

\subsection{Stability of GREC}
\label{sec:stab}
The efficiency of GREC can be understood in two ways. First, the GREC algorithm ideologically stems from the PEC and ZNE methods, and GREC's summation of the randomized circuits, trained on the classical data, can be viewed as a peculiar form of the probabilistic error cancellations or as a way to take the zero-noise limit. Second, since the classically available data is the basic element of the GREC algorithm, it can be interpreted as an extrapolation of classical calculations to the large range of parameters with the support of quantum computations. Here we follow the later point of view. It allows us to apply recent developments in the extrapolation theory \cite{Stable} and to estimate the stability of the GREC algorithm with respect to the number of additionally randomized circuits and the numerical accuracy of the classical data.

To start with the analysis, we represent noisy curves measured on the $N_\R$ randomized circuits as
\begin{equation}
  \langle \A(\lambda) \rangle^{(r)} = \langle \A(\lambda) \rangle + \delta\langle \A(\lambda) \rangle^{(r)}\,,~~~r = 1..N_\R\,.
\end{equation}
For the following we assume that functions $\delta\langle \A(\lambda) \rangle^{(r)}$ are analytic in the complex plane inside the Bernstein ellipse with foci at $\pm 1$ and semiminor and semimajor axis lengths summing to $\rho > 1$, denoted by $E_\rho$, and bounded in $E_\rho$ as $\delta\langle \A(\lambda) \rangle^{(r)} \leq Q$ for $\lambda \in E_\rho$, $Q < \infty$.
The constrain $\sum_{r = 1}^{N_\R} \eta_r = 1$ implies that
\begin{equation}
  \langle \A(\lambda) \rangle^{\rm mitigated} = \langle \A(\lambda) \rangle + \sum_{r = 1}^{N_\R} \eta_r \delta\langle \A(\lambda) \rangle^{(r)} + \eta_0\,.
\label{mitig}
\end{equation}
Ideally, $\sum_{r = 1}^{N_\R} \eta_r \delta\langle \A(\lambda) \rangle^{(r)} + \eta_0 \approx 0$. Let us expand each error term $\delta\langle \A(\lambda) \rangle^{(r)}$ in the Chebyshev polynomial basis
\begin{equation}
  \delta\langle \A(\lambda) \rangle^{(r)} = \sum_{n = 0}^{N_\R} c_n T_n(\lambda) + R^{(r)}(N_\R, \lambda)\,.
\label{trunc}
\end{equation}
For the functions $\delta\langle \A(\lambda) \rangle^{(r)}$ satisfying assumptions above the reminder term can be bounded as \cite{Stable},
\begin{eqnarray}
  \big|R^{(r)}(N_\R, \lambda)\big| <  \frac{2 Q \rho^{-N_\R}}{(\rho - 1)}\,,~~~\lambda \in \K\,.
\end{eqnarray}
The total error arising due to the truncation's \eqref{trunc} in \eqref{mitig} is given by $\bar\epsilon = \frac{2 Q \rho^{-N_\R}}{(\rho - 1)} \sum_{r = 1}^{N_\R} |\eta_r|$. In practical computations one can always restrict parameters $\eta_r$ to satisfy $|\eta_r| < 1$, then the truncation error estimate simplifies as $\bar\epsilon = \frac{2 Q \rho^{-N_\R}}{(\rho - 1)} N_\R$. 
The optimization \eqref{opt} together with the construction \eqref{constr} is equivalent to finding the extrapolation of one of the error terms from the sum $\sum_{r = 1}^{N_\R} \eta_r \delta\langle \A(\lambda) \rangle^{(r)}$, let say with $r = 1$. Let also $|\eta_1| = \max_{r = 1..N_\R}\{|\eta_r|\}$, we have
\begin{equation}
  \delta\langle \A(\lambda) \rangle^{(1)}_{\rm extrap} = \sum_{n = 0}^{N_\R} b_n T_n(\lambda) + \hat\epsilon\,,~~~\hat\epsilon \leq \frac{\bar\epsilon}{|\eta_1|}.
\label{A0S}
\end{equation}
The quality of such extrapolation strongly depends on the analytic properties of the function $\delta\langle \A(\lambda) \rangle^{(1)}$. Assume the function $\delta\langle \A(\lambda) \rangle^{(1)}$ can be evaluated on the classical computer in $N_\M = 4 N_\R^2$ equally distributed points in the region $\K = [-1, 1]$ with the precision $\epsilon > 0$. Then, $\delta\langle \A(\lambda) \rangle^{(1)}$ can be stably extrapolated to $\lambda \in \I_\rho [1, (\rho + \rho^{-1})/2)$ by the truncated series \eqref{A0S} with coefficients determined by the least square method with the error given by \cite{Stable}
\begin{equation}
 \Big|\delta\langle \A(\lambda) \rangle^{(1)}_{\rm extrap} - \delta\langle \A(\lambda) \rangle^{(1)}\Big| \leq C_{\rho, \epsilon + \hat\epsilon} \frac{Q}{1 - {\textgoth r}(\lambda)} \Big(\frac{\epsilon + \hat\epsilon}{Q}\Big)^{\alpha(\lambda)}\,
\label{bound}
\end{equation}
where
\begin{eqnarray}
  {\textgoth r}(\lambda) = \frac{\lambda + \sqrt{\lambda^2 -1}}{\rho}\,,~~~~\alpha(\lambda) = -\frac{\log {\textgoth r}(\lambda)}{\log\rho}
\end{eqnarray}
and $C_{\rho, \epsilon + \hat\epsilon}$ is a constant which depends only polylogarithmically on $1/(\epsilon + \hat\epsilon)$.

The extrapolation error of \eqref{A0S} is proportional to the fractional power of $(\epsilon + \hat\epsilon)$ for $\lambda$ inside the range $\I_\rho$. When $\lambda = 1$, the bound \eqref{bound} is proportional to the sum of errors $(\epsilon + \hat\epsilon)$. When $\lambda$ approaches the right end of the range, $(\rho + \rho^{-1})/2$, the power of $(\epsilon + \hat\epsilon)$, $\alpha(\lambda) \rightarrow 0$, at the same time the bound \eqref{bound} diverge due to the prefactor $\frac{Q}{1 - {\textgoth r}(\lambda)}$, since in this case, ${\textgoth r}(\lambda) \rightarrow 1$. This bound is almost tight, meaning it cannot be meaningfully improved by any other linear or nonlinear procedure of constructing the extrapolant \cite{Stable}.

Therefore, we conclude that independently from the cancellations of quantum errors supposed in the expansions of the type \eqref{decomp}, the GREC algorithm should be efficient for the intermediate values of $\lambda$ being sufficiently far away from the right end of the range $\I_\rho$. In case of significant cancellations of quantum errors in \eqref{constr}, the resulting approximation of an observable can be obtained with much better precision than in accordance with the almost tight bound \eqref{bound}.

\section{Anti-ferromagnetic Ising model in the transverse field}
The Hamiltonian of the anti-ferromagnetic Ising chain of $n$ spins in the transverse field is given by
\begin{equation}
\mathcal{H}=\sum_{i=1}^{n}\sigma_{i}^{x}\sigma_{i+1}^{x} + \sigma_{1}^{y}\sigma_{2}^{z}\cdots\sigma_{n-1}^{z}\sigma_{n}^{y} +\lambda\sum_{i=1}^{n}\sigma_{i}^{z}\,,
\label{H}
\end{equation}
where $\sigma^x$, $\sigma^y$, $\sigma^z$ are Pauli matrices and $\lambda$ is the transverse field strength. 
The Hamiltonian \eqref{H} differs from the conventional one by the presence of the second term negligible in the large $n$ limit and added in order to cancel the periodic boundary term to make the system exactly solvable as in the case of an infinite chain  \cite{ExactIsing}. The diagonalization of the Hamiltonian \eqref{H} is achieved by applying the unitary transformation $\tilde U$ composed by the consequent application of the Jordan-Wigner, Fourier, and Bogoliubov transformations.
The transformation $\tilde U$ can be implemented as a quantum circuit \cite{ExactIsing}, see Supplemental Material. The exact solvability anti-ferromagnetic Ising chain together with the possibility to express the solution as a quantum circuit provides an ideal setup for testing the GREC quantum error mitigation. The central quantity for our following numerical computations is the average ground state magnetization, which can be directly measured by running the quantum circuits or computed analytically as
\begin{equation}
\langle \sigma_z(\lambda) \rangle = 
  \begin{cases} 
      \frac{\lambda}{2\sqrt{1+\lambda^2}} & \lambda < 1 \\
      \frac{1}{2} + \frac{\lambda}{2\sqrt{1+\lambda^2}} & \lambda > 1\,.
  \end{cases}
\label{exact}
\end{equation}
The piecewise structure of the analytic solution reflects the phase transition occurring at the external magnetic field with the strength $\lambda = 1$ in the case of the infinite chain. Two branches of \eqref{exact} correspond to two slightly different quantum circuits, so for simplicity of presentation, in our numerical analysis, we focus only on $\lambda > 1$, which is a more challenging range for quantum computations than $0 < \lambda < 1$, \cite{ExactIsing}.

\section{Results of experiments}
\label{ExRes}
We study the performance of the GREC error mitigation by applying it to the magnetization of the $n = 4$ spins Ising model in the transverse magnetic field of the strength $\lambda \in \Lambda = [1, 3.5]$. For our investigations, we use the data obtained by the classical simulations of the \textit{ibmq\_manila} quantum computer and by the real quantum experience with \textit{ibmq\_manila}, \cite{IBMQ}. 
We choose ranges $\K_1 = [1, 2]$ and $\K_2 = [2.5, 3.5]$ as accessible for the classical computations and split them into the training and validating parts $\K_i = \T_i \cup \V_i$, $i=\{1, 2\}$, so $\T_1 = [1.5, 2]$, $\V_1 = [1, 1.5]$ and $\T_2 = [2.5, 3]$, $\V_2 = [3, 3.5]$. While the training sets are used for the optimization \eqref{opt}, the performance of GREC on the validation sets is utilized for fixing the reasonable number of the randomized curves $N_\R$ and corresponding value of $\Delta$. Additionally, the validation may be also used for selecting the most optimal configuration of the randomizing gates $\G_l(\vec{\theta_l})$.
In computations based on the classical simulations training and validating ranges contain $10$ points. In the real quantum computing experience, the fixing of parameters $N_\R$ and $\Delta$ based on the validation is not applied due to the deficiency of computational time. The values of $N_\R$ and $\Delta$ are taken as classical simulations and the number of points in ranges $\V_1$ and $\V_2$ is reduced to $5$.
In all our computations we use $N_\R = 9$ and $\Delta = 0.1$. 
We modify the original circuit by completing each single-qubit gate by the consequent variational $U(\vec\theta)$ gate, where the $3$-component vector $\vec{\theta}$ provides its general parameterization. This results in $N_\G = 10$ auxiliary gates for the whole circuit. 
We didn't observe any significant difference in taking parameters of the randomized circuits from the range $[-\Delta, \Delta]$ or from $[0, \Delta]$, and in the following examples we use $\vec{\theta}_l$ uniformly distributed in $[0, \Delta]$.

\subsection{Experiments on quantum simulator}
\begin{figure}[h]
\includegraphics[width=0.45
\textwidth]{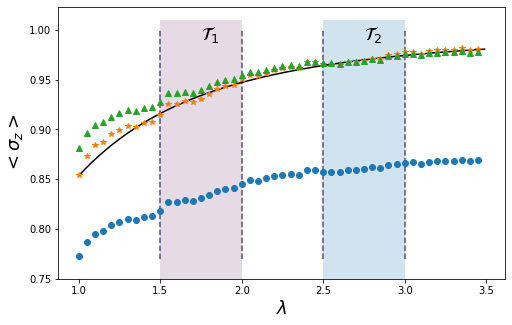}
\caption{\small The baseline of GREC. The solid black line represents the exact solution. Blue circles represent the ground state magnetization obtained in initial classical simulations. Orange stars and green triangles show the results with mitigated errors and correspond to regression coefficients $\eta_0$, $\eta_1$ obtained on ranges $\T_1 = [1.5, 2]$ and $\T_2 = [2.5, 3]$ respectively.}
\label{fig:ppf}
\end{figure}

We present the application of the baseline version of the GREC algorithm -- a two-parameter linear fitting -- in Fig. \ref{fig:ppf}. This fitting doesn't take into account the dependence of quantum errors on the parameter $\lambda$ and can be considered only as an 'optimal in average' for $\lambda \in \Lambda = [1, 3.5]$. Despite the reasonable quality of the error mitigation, it is seen from Fig. \ref{fig:ppf} that the results are different for ranges $\K_1$ and $\K_2$. 

\begin{figure}[h]
\centering
\includegraphics[width=0.45\textwidth]{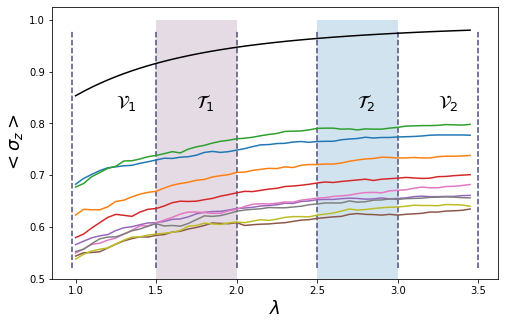}
\caption{\small 9 randomized curves sampled with $\Delta = 0.1$, classical simulations of \textit{ibmq\_manila}. The upper black line shows the exact solution, other color lines represent magnetization's computed using randomized circuits. The training and validating ranges are indicated by $\T_1$, $\V_1$ and $\T_2$, $\V_2$.}
\label{fig:rand}
\end{figure}

\begin{figure}[h]
\centering
\includegraphics[width=0.45\textwidth]{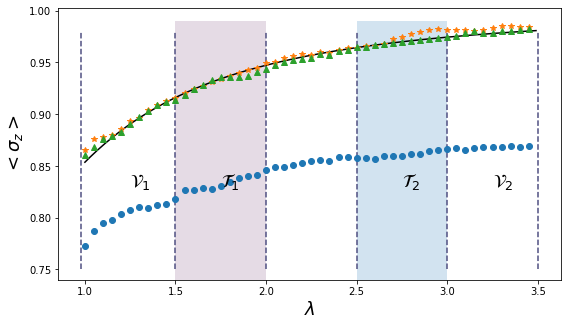}
\caption{\small GREC error mitigation applied to simulations of \textit{ibmq\_manila}. The solid black line -- exact solution. Blue circles -- initially obtained magnetization. Orange stars and green triangles represent the results with mitigated errors and correspond to training and validating ranges $\T_1 = [1.5, 2]$, $\V_1 = [1, 1.5]$ and $\T_2 = [2.5, 3]$, $\V_2 = [3, 3.5]$ respectively.}
\label{fig:GREC}
\end{figure}

We demonstrate the work of the full GREC algorithm by presenting corresponding randomized magnetization curves in Fig. \ref{fig:rand} and mitigated magnetization's obtained for classically accessible ranges $\K_1$ and $K_2$ with the training and validation on $\T_1$, $\T_2$, and $\V_1$, $\V_2$ respectively in Fig. \ref{fig:GREC}.

\subsection{Experiments on real quantum processor}
\begin{figure}[h]
\centering
\includegraphics[width=0.45\textwidth]{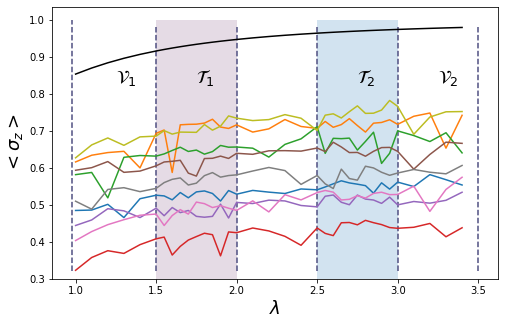}
\caption{\small 9 randomized curves obtained by the real quantum experience with \textit{ibmq\_manila}, $\Delta = 0.1$. The upper black line shows the exact solution, other color lines represent magnetization's computed using randomized circuits. The training and validating ranges are indicated by $\T_1$, $\V_1$ and $\T_2$, $\V_2$.}
\label{fig:GREC_QUANTUM_RAND}
\end{figure}

\begin{figure}[h]
\centering
\includegraphics[width=0.45\textwidth]{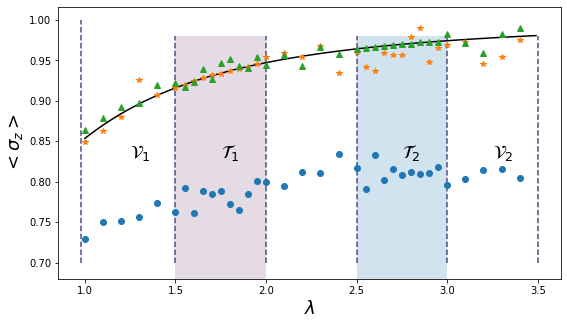}
\caption{\small GREC error mitigation applied to the real quantum experience with \textit{ibmq\_manila}. The solid black line -- exact solution. Blue circles -- initially obtained magnetization. Orange stars and green triangles represent the results with mitigated errors and correspond to training and validating ranges $\T_1 = [1.5, 2]$, $\V_1 = [1, 1.5]$ and $\T_2 = [2.5, 3]$, $\V_2 = [3, 3.5]$ respectively.}
\label{fig:GREC_QUANTUM}
\end{figure}

For demonstrating the efficiency of the GREC algorithm with the real quantum experience, we focus only on the full version of GREC. The computations on the \textit{ibmq\_manila} were performed from... to... . In Fig. \ref{fig:GREC_QUANTUM_RAND} we show $N_\R = 9$ randomized curves and in Fig. \ref{fig:GREC_QUANTUM} we present mitigated magnetization's trained on ranges $\T_1$ and $T_2$.

\section{Concluding remarks}
In this Letter, we studied the quantum error mitigation for the parametric quantum circuits, which can be efficiently simulated on classical computers for some ranges of their parameters. The utilization of such additional information significantly differs the GREC algorithm from other methods and allows one to achieve better suppression of quantum errors, see Supplemental Material for the simultaneous presentation of the GREC and ZNE computations.

Interpreting the work of GREC as an analytic continuation of the function correcting quantum errors from the range of parameters, where the classical computations are manageable to the rest set of parameters, we have shown that the error of the GREC algorithm is proportional to the fractional power of the sum of the error of the classical computation and the error caused by the finite amount of the randomized quantum circuits. This bound is almost tight when the summations of randomized quantum circuits in the GREC algorithm don't lead to the cancellation of quantum errors. In the considered example of the $n=4$ Ising spin chain in the transverse magnetic field, the application of GREC leads to reasonable results for the magnetization curve even with small amounts of training points and randomized circuits. This indicates the significant cancellation of quantum errors in these computations.

The ultimate applications of the GREC method should include such important problems as quantum computations in lattice QCD and Hubbard model at finite chemical potentials, where the non-physical regions with purely imaginary chemical potentials are accessible by the classical Monte Carlo simulations. For the first steps towards these ambitious goals, one may investigate the application of the GREC algorithm to the lattice $\phi^4$ or Schwinger models, which have similar properties regarding the \textit{sign problem} within the conventional Monte Carlo methods and also have \textit{sign problem}-free formulations in terms of dual variables and on quantum computers \cite{phi4SP, phi4Q, SchwingerSP, SchwingerSP2, SchwingerQ, IBMgauge}.

\begin{acknowledgments}
We acknowledge the use of IBM Quantum services for this work. The views expressed are those of the authors, and do not reflect the official policy or position of IBM or the IBM Quantum team.
\end{acknowledgments}

\onecolumngrid
\section*{SUPPLEMENTAL MATERIAL}

\subsection{Comparison with ZNE}
\begin{figure}[h]
\centering
\includegraphics[width=0.7\textwidth]{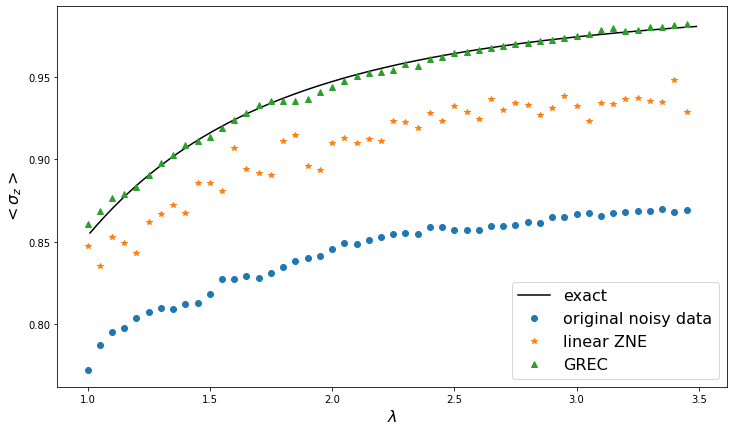}
\caption{\small Comparison of the GREC and linear ZNE quantum error mitigations.}
\label{fig:GRECvsZNE}
\end{figure}

\begin{figure}[h]
\centering
\includegraphics[width=0.5\textwidth]{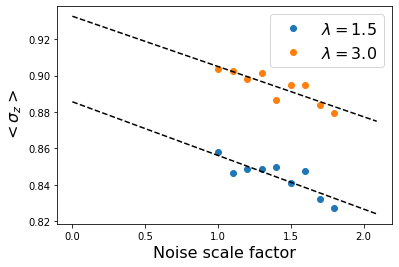}
\caption{\small Linear zero noise extrapolation of the ground state magnetization for $\lambda = 1.5$ and $\lambda = 3.0$.}
\label{fig:ZNE_extr}
\end{figure}

The zero-noise extrapolation is a general method that can be applied when the underlying noise model is unknown. However, it can be sensitive to extrapolation errors. Therefore, one has to take care about choosing the appropriate set of scale factors, extrapolation method, and the noise-scaling method, which in most cases are not known. In our reference computations, we used the linear ZNE implemented in the Mitiq open-source library \cite{Mitiq} with the set of $9$ scaling factors equally distributed in  $[1.0, 1.9]$. This corresponds to the $9$ quantum computations per one point of the magnetization curve -- exactly as in our computations with the GREC method. We present the results obtained by the linear ZNE and by the GREC method with the training region $\K = [1.5, 2.0]$ in Fig. \ref{fig:GRECvsZNE}. In Fig. \ref{fig:ZNE_extr} we show two examples of the linear zero noise extrapolation corresponding to points of the magnetization curve with $\lambda = 1.5$ and $\lambda = 3.0$.

\subsection{Quantum circuits}
\begin{figure}[h]
\centering
\includegraphics[width=\textwidth,trim={6cm 3cm 6cm 3cm},clip]{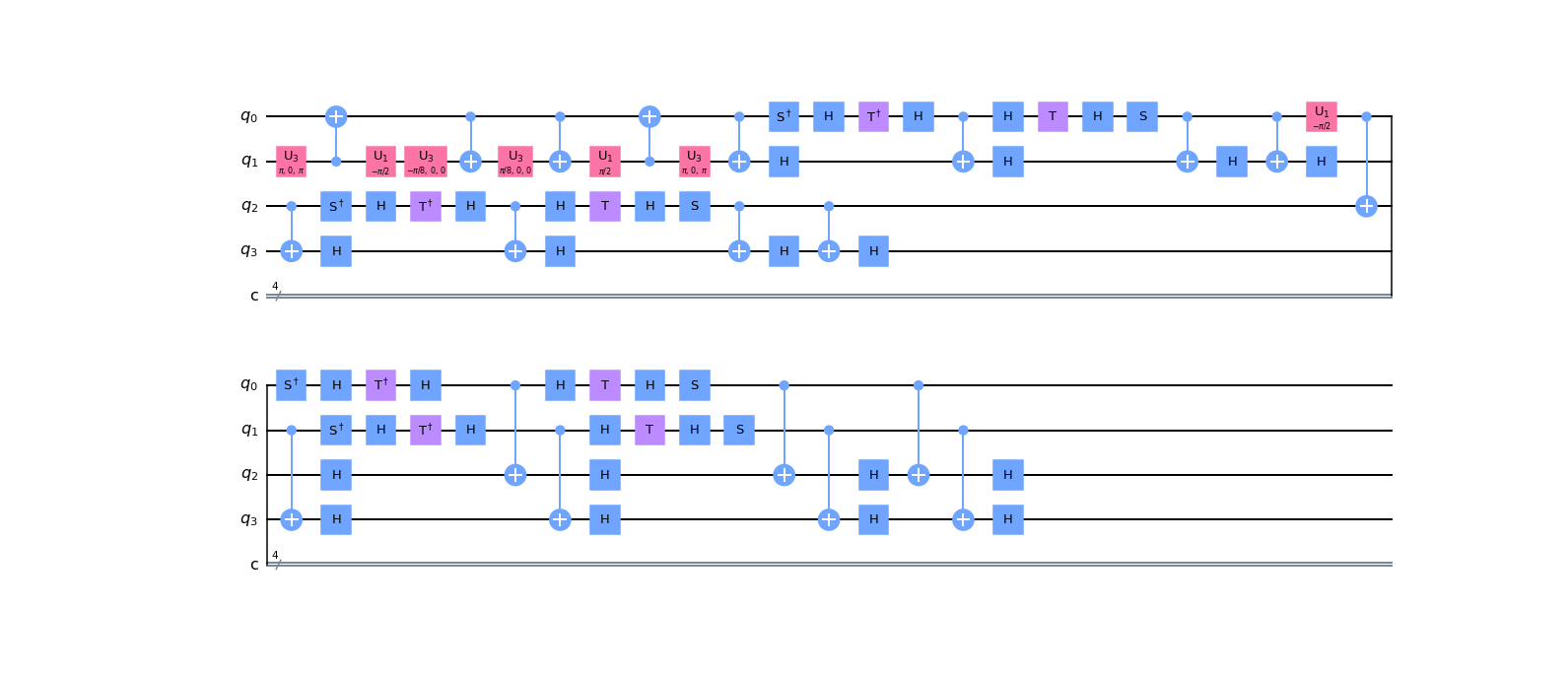}
\caption{\small Quantum circuit for $n=4$ spin Ising model with numerical values of the gate's parameters corresponding to $\lambda = 1$.}
\label{fig:qc_origin}
\end{figure}

\begin{figure}[h]
\centering
\includegraphics[width=0.9\textwidth]{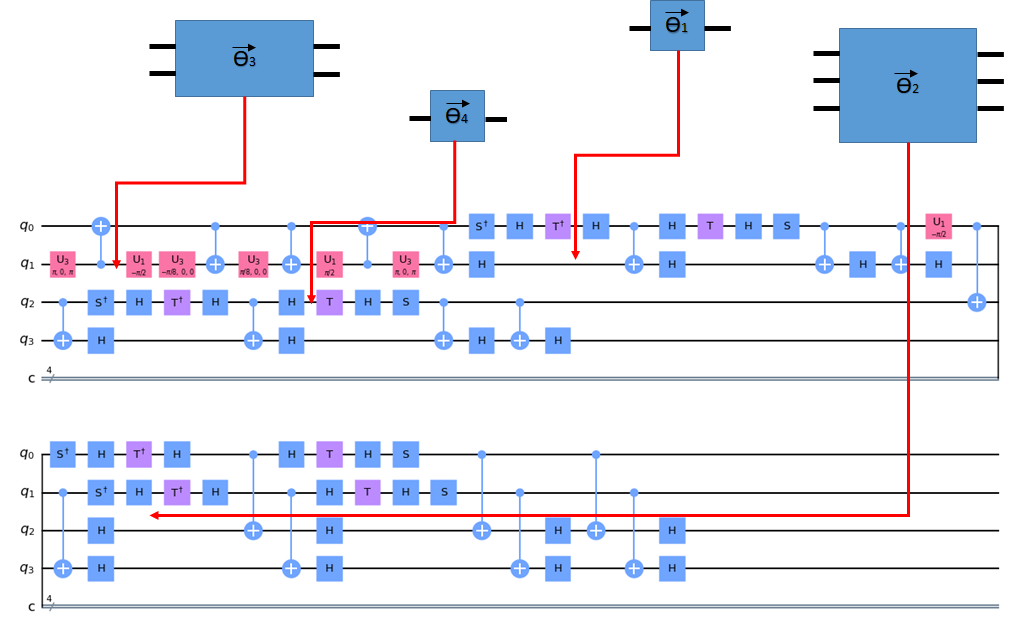}
\caption{\small Schematic representation of random insertions of four parametric gates.}
\label{fig:gen_rand}
\end{figure}

In Fig. \ref{fig:qc_origin} we present the quantum circuit for $n=4$ spin Ising model valid for $\lambda \geq 1$.
The general procedure of the circuit randomization can be performed by random insertions of parametric gates acting on different amount of qubits in the original quantum circuit, we show it schematically in Fig. \ref{fig:gen_rand}.
To prepare the randomized circuits used in our GREC computations for the $n=4$ spin Ising model, we have changed each different one-qubit gate by an 'equipped one', as
\begin{eqnarray}
U_1(-\pi/2) \rightarrow U_1(-\pi/2) U(\vec\theta_1)\,,~~~U_1(\pi/2) \rightarrow U_1(\pi/2) U(\vec\theta_2)\,,~~~H \rightarrow H U(\vec\theta_3)\,,\dots,
\end{eqnarray}
resulting in introducing of $10$ random vectors $\vec\theta_l$ corresponding to $30$ real valued parameters. We have choose each of them randomly from the region $[0, 0.1]$ for each of $9$ randomized circuits.


\begin{thebibliography}{2}
\bibitem [{a()}]{a}%
  \BibitemOpen
  \bibinfo {note} {In principle, one can include a non-randomized measurements
  ${{\protect \mathcal A}}_\Lambda \equiv {{\protect \mathcal A}}_\Lambda
  ^{(0)}$ into this set, however we didn't find any remarkable advantage of
  it.}\BibitemShut {Stop}%
\bibitem [{b()}]{b}%
  \BibitemOpen
  \bibinfo {note} {This constrain is introduced to make the derivations of
  Section \ref {sec:stab} more transparent, and it was omitted in the baseline
  version of the GREC algorithm.}\BibitemShut {Stop}%
\end{thebibliography}
\end{document}